\documentclass[12pt,a4paper]{article}
\usepackage{latexsym}
\usepackage{epsfig}
\usepackage{url}
\usepackage{calc}

\newcommand{\mfantom}[1]{\left.\hspace{#1}\right.}
\newcommand\eq[1] {(\ref{#1})}
\newcommand{\bfm}[1]{\mbox{\boldmath ${#1}$}}

\newcommand{\bequ}{\begin{equation}}
\newcommand{\eequ}[1]{\label{#1}\end{equation}}

\newcommand{\sign}{\mbox{sign}}

\newcounter{hours}\newcounter{minutes}
\newcommand{\printtime}{\setcounter{hours}{\time/60}
  \setcounter{minutes}{\time-\value{hours}*60}
  \thehours :\theminutes}

\newcommand\cond {{\rm cond}}
\newcommand\varkappa {\kappa}

\begin{document}

\title{Estimation of the linear transient growth of
perturbations of cellular flames}
\author{V. Karlin}
\date{}
\maketitle

\thispagestyle{empty}

\begin{center}
Centre for Research in Fire and Explosion Studies\\
University of Central Lancashire,
Preston PR1 2HE, UK\\
Email: VKarlin@uclan.ac.uk
\end{center}

\centerline{
Compiled on \today\hspace{1pt} at\hspace{-8pt} \printtime\hspace{1pt}
}

\vspace{5mm}
\begin{abstract}
In this work we estimate rates of the linear transient
growth of the perturbations of cellular flames governed
by the Sivashinsky equation. The possibility
and significance of such a growth was indicated earlier
in both computational and analytical investigations.
Numerical investigation of the norm of the resolvent of
the linear operator associated with the Sivashinsky
equation linearized in a neighbourhood of the steady
coalescent pole solution was undertaken. The results are
presented in the form of the pseudospectra and the
lower bound of possible transient amplification.
This amplification is strong enough to make the
round-off errors visible in the numerical simulations
in the form of small cusps
appearing on the flame surface randomly in time.
Performance of available numerical approaches was
compared to each other and the results are checked
versus directly calculated norms of the evolution
operator.
\end{abstract}

\vspace{3mm}\noindent
{\bf Key words:} nonnormal operator, pseudospectra,
nonmodal amplification,
hydrodynamic flame instability, cellular flames

\vspace{3mm}\noindent
{\bf PACS 2003:} 47.70.Fw, 47.20.Ma, 47.20.Ky, 47.54.+r, 02.60.Nm

\vspace{3mm}\noindent
{\bf ACM computing classification system 1998:} J.2, G.1.9, G.1.3, G.1.0

\vspace{3mm}\noindent
{\bf AMS subject classification 2000:} 80A25, 76E15, 76E17, 35S10, 65G50

\vspace{3mm}\noindent
{\bf Abbreviated title:} Transient growth of perturbations of
cellular flames

\newpage

\section{Introduction}

\begin{figure}[ht]

\centerline{\epsfig{figure=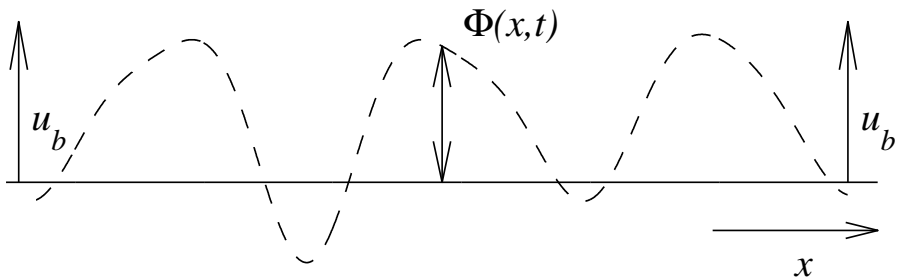,height=40mm,width=120mm}}

\caption{Perturbation (dashed line) of the plane flame front
(solid line) moving at a speed $u_{b}$.}
\label{pert}
\end{figure}

Sivashinsky's equation
\bequ
\frac{\partial\Phi}{\partial t}
-\frac{1}{2}\left(\frac{\partial\Phi}{\partial x}\right)^{2}
=\frac{\partial^{2}\Phi}{\partial x^{2}}
+\frac{\gamma}{2}\frac{\partial{\cal H}[\Phi]}{\partial x},\qquad
x\in\bfm R,\qquad t> 0,
\eequ{1a}
governs evolution of the perturbation $\Phi(x,t)$
of the plane flame front moving in the direction
orthogonal to the $x$-axis with the laminar flame
speed $u_{b}$, see Fig. \ref{pert}. Here space coordinates
are measured
in units of the flame front width $\delta_{th}$,
time is in units of $\delta_{th}/u_{b}$, and
${\cal H}[\Phi]=\pi^{-1}\int\limits_{-\infty}^{\infty}
(x-y)^{-1}\Phi(y,t)dy$ is the Hilbert transform.

The equation was obtained in \cite{Sivashinsky77a}
considering the flame front as a surface separating combustible
mixture of density $\rho_{u}$ and burnt gases of
density $\rho_{b}$. Assumptions of the low expansion rate
$\rho_{b}/\rho_{u}\approx 1$ and small flame surface
gradient $|\nabla\Phi|\ll 1$ were also used in order to
justify the appearance of the nonlinearity in \eq{1a},
where the parameter $\gamma=1-\rho_{b}/\rho_{u}$. 

A wide class of periodic solutions to \eq{1a} was obtained
in \cite{Thual-Frisch-Henon85} by using the pole decomposition
technique. Namely, it was shown that
$$
\Phi(x,t)=2\pi NL^{-1}\left(\gamma
-4\pi NL^{-1}\right)t
$$
\bequ
+2\sum\limits_{n=1}^{N}
\ln\left|\cosh\left[2\pi b_{n}(t)/L\right]
-\cos\left\{2\pi[x-a_{n}(t)]/L\right\}\right|
\eequ{1b}
is an $L$-periodic solution to \eq{1a} if
$$
\frac{da_{n}}{dt}=
-\frac{2\pi}{L}\sum\limits_{m=1}^{N}\mfantom{-6pt}^{'}
\left\{\frac{\sin[2\pi(a_{n}-a_{m})/L]}
{\cosh[2\pi(b_{n}-b_{m})/L]-\cos[2\pi(a_{n}-a_{m})/L]}\right.
$$
\bequ
+\left.\frac{\sin[2\pi(a_{n}-a_{m})/L]}
{\cosh[2\pi(b_{n}+b_{m})/L]-\cos[2\pi(a_{n}-a_{m})/L]}\right\},
\eequ{1c_a}

$$
\frac{db_{n}}{dt}= 
2\pi L^{-1}\coth\left(2\pi b_{n}/L\right)
-(\gamma/2)\sign b_{n}
$$
$$
+\frac{2\pi}{L}\sum\limits_{m=1}^{N}\mfantom{-6pt}^{'}
\left\{\frac{\sinh[2\pi(b_{n}-b_{m})/L]}
{\cosh[2\pi(b_{n}-b_{m})/L]-\cos[2\pi(a_{n}-a_{m})/L]}\right.
$$
\bequ
+\left.\frac{\sinh[2\pi(b_{n}+b_{m})/L]}
{\cosh[2\pi(b_{n}+b_{m})/L]-\cos[2\pi(a_{n}-a_{m})/L]}\right\}.
\eequ{1c_b}
Here $N$ is an arbitrary positive integer and prime
in the symbol of summation means $m\ne n$. Pairs of real
numbers $(a_{n},b_{n})$, $n=1,\ldots,N$
are called poles and, correspondingly, function \eq{1b} is
called $N$-pole\footnote{Strictly speaking, $N$ is the
number of complex conjugated pairs of poles $a_{n}\pm ib_{n}$.
However, we follow the tradition and keep this natural definition,
as only real solutions are of interest.}
solution to \eq{1a}. It is also convenient to consider
$\Phi(x,t)\equiv {\rm const}$ as a $0$-pole solution to \eq{1a}.

\begin{figure}[ht]

\centerline{\epsfig{figure=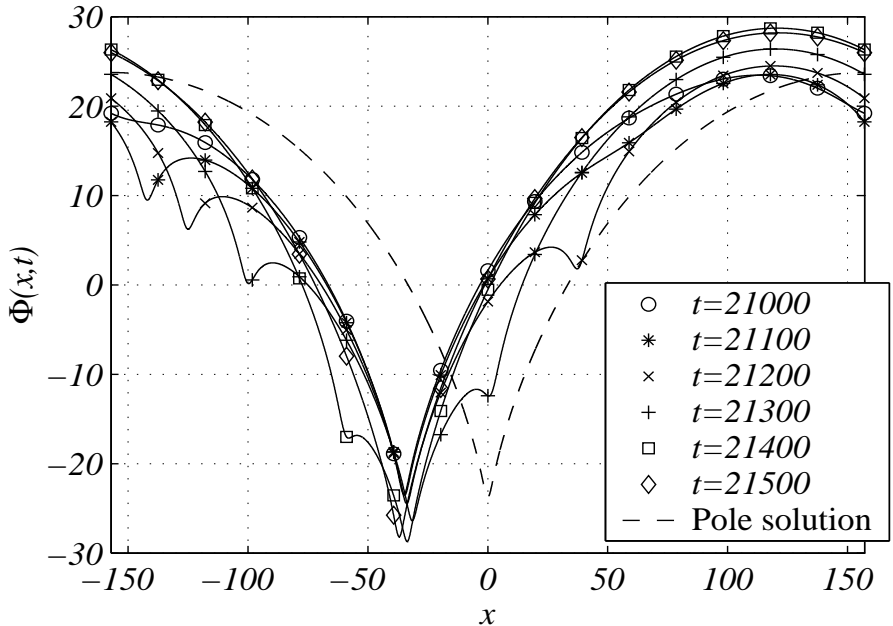,height=85mm,width=120mm}}

\caption{Steady coalescent $N$-pole solutions to the Sivashinsky
equation. Here $\gamma=0.8$ and $L=100\pi$ ($N_{L}=10$).
Graphs have been shifted vertically in order to
get $\Phi_{N}(\pm L/2)=0$.}
\label{micro_cusps}
\end{figure}

If all the poles in \eq{1c_a}, \eq{1c_b} are steady and
$a_{n}=a\in\bfm R$
for $n=1,\ldots,N$, then, \eq{1b} is called a steady
coalescent $N$-pole solution. Solutions of the latter type,
denoted here as $\Phi_{N}(x)$ and illustrated in Fig.
\ref{micro_cusps}, have been found to be the strongest
attractors of \eq{1a} and the period $L$ preferred by
\eq{1a} has appeared to coincide with the size of the whole
computational domain which we therefore denote as $[-L/2,L/2]$,
see e.g. \cite{Rahibe-Aubry-Sivashinsky96}.

It was shown, see for example \cite{Rahibe-Aubry-Sivashinsky98},
that for a given period $L$ the number of poles in steady
coalescent pole solution \eq{1b} may not exceed
$N_{L}={\rm ceil}(\gamma L/8\pi+1/2)-1$, where ${\rm ceil}(x)$
is the smallest integer greater or equal to $x$. Direct
numerical simulations have revealed, in turn, that for
sufficiently small values of $L<L_{c}$ the preferred number
of poles is equal $N_{L}$. This observation was explained
in \cite{Vaynblat-Matalon00a} by means of the eigenvalue
analysis of \eq{1a} linearized in a neighbourhood of the
steady coalescent pole solutions. The analysis has
indicated that for any $L>0$ the steady coalescent
 $N_{L}$-pole solution is the only steady coalescent
 $N$-pole solution to \eq{1a} with all the
eigenvalues located in the left half of the complex plane.
Strictly speaking, \cite{Vaynblat-Matalon00a} does not
provide a solid proof that their set of eigenvalues is
complete and in this paper we explain why the comparison
with the direct numerical calculation of the spectra, used
in \cite{Vaynblat-Matalon00a}, cannot justify the completeness,
in particular for large enough $L$.

Surprisingly, for larger computational domains $L>L_{c}$,
numerical solutions to \eq{1a} do not stabilize to any
steady coalescent $N$-pole solution at all.
Instead, being essentially
nonsteady, they remain very closely to the steady coalescent
$N_{L}$-pole solution, developing on the surface of the flame
front small cusps randomly in time, see e.g. \cite{Karlin02}.
With time these small cusps move towards the trough of the
flame front profile and disappear in it as this is shown in
Fig. \ref{micro_cusps}.

The high sensitivity of pole solutions to certain
perturbations was suggested in \cite{Joulin89b} as an
explanation of the cardinal change in the 
behaviour of numerical solutions to \eq{1a}
which takes place for $L=L_{c}$.
The argument of \cite{Joulin89b} was based on a particular
asymptotic solution to an approximation of the Sivashinsky
equation linearized in a neighbourhood of the steady
coalescent $N$-pole solution. In the
following works, see e.g. \cite{Cambray-Joulin92},
the approach has been developed further and a model equation
with stochastic right hand side, explicitly representing
the noise, has been proposed and investigated.
Sensitivity of Sivashinsky equation to
the noise has also been studied in
\cite{Olami-Galanti-Kupervasser-Procaccia97}, and an
estimation of dependence between $L_{c}$ and the
amplitude of noise
in the form of the the round-off errors has been obtained in
\cite{Karlin02} in a series of direct numerical
simulations.

Similar insufficiency of the eigenvalue analysis to interpret
time dependent behaviour of asymptotically stable
systems is also known from problems of classic
hydrodynamics, such as Poiseuille and Hagen-Poiseuille
flows \cite{Trefethen-Trefethen-Reddy-Driscoll93}.
The failure of the spectral
analysis in these problems was linked to the nonorthogonality
of the eigenfunctions of the associated linearized operators
and was explained by the estimation of possible transient
growth of perturbations \cite{Boberg-Brosa88},
\cite{Reddy-Schmid-Henningson93}. A convenient tool to
estimate possible transient growth of solutions governed
by nonnormal operators were developed during the
last decade in the form of the pseudospectra
\cite{Trefethen97}. Corresponding numerical techniques
have been reviewed in \cite{Trefethen99}.

In this work we estimate rates of the linear transient
growth of the perturbations of the steady coalescent
$N_{L}$-pole solutions to the Sivashinsky equation.
In Section 2 we linearize the equation in a neighbourhood
of the steady coalescent pole solution.
In Section 3, results of direct computations of the
pseudospectra of the linear operator are presented.
Also, a comparison of performance of available
numerical techniques is given.
Estimation of the rates of growth in terms
of Kreiss constants, norms of the $C_{0}$-semigroup
and condition numbers, are presented in Section 4.
We conclude with a discussion and a summary of results
in Section 5.

\section{Linearized Sivashinsky equation}\label{nillspace}

Substituting
$\Phi(x,t)=\Phi_{N}(x,t)+\phi(x,t)$ into \eq{1a}
and neglecting terms which are nonlinear in $\phi(x,t)$,
one obtains
\bequ
\frac{d\phi}{dt}={\cal A}_{N}\phi,\qquad t>0,
\eequ{2_1c}
where operator ${\cal A}_{N}$ is defind by the following
integro-differential expression
\bequ
{\cal A}_{N}u
=\Psi_{N}\frac{\partial u}{\partial x}
+\frac{\partial^{2} u}{\partial x^{2}}
+\frac{\gamma}{2}\frac{\partial{\cal H}[u]}{\partial x},\qquad
x\in\bfm R,
\eequ{2_1d}
on sufficiently smooth $L$-periodic functions with the square
integrable on $[-L/2,L/2]$. Here
\bequ
\Psi_{N}=\frac{\partial\Phi_{N}}{\partial x}
=\frac{4\pi}{L}\sum\limits_{n=1}^{N}
\frac{\sin[2\pi(x-a)/L]}{\cosh(2\pi b_{n}/L)-\cos[2\pi(x-a)/L]},
\eequ{2_1b}
$L>0$ and $a\in\bfm R$ are real parameters, and the set
$b_{n}$, $n=1,\ldots,N$ is the steady solution to
\eq{1c_b}.

The adjoint integro-differential expression is
\bequ
{\cal A}_{N}^{*}u
=-\frac{\partial\Psi_{N}u}{\partial x}
+\frac{\partial^{2} u}{\partial x^{2}}
+\frac{\gamma}{2}\frac{\partial{\cal H}[u]}{\partial x},\qquad
x\in\bfm R.
\eequ{2_1e}
Hence, ${\cal A}_{N}{\cal A}_{N}^{*}$
$\ne{\cal A}_{N}^{*}{\cal A}_{N}$ for $N>0$,
and operator ${\cal A}_{N}$ is nonnormal. 
However, for $0$-pole solution the first term in the
right hand side of \eq{2_1d} disappears and operator
${\cal A}_{N}$ is normal. In this sense we can say
that it is the nonlinearity of the Sivashinsky equation
what makes its associated linearized operator
${\cal A}_{N}$ nonnormal.

If \eq{2_1c}, \eq{2_1d} is differentiated by $x$, then
the resulting equation for $\psi(x)=d\phi/dx$ is
$\partial\psi/\partial t={\cal A^{\prime}}_{N}\psi$,
where
$$
{\cal A^{\prime}}_{N}\psi
=\frac{\partial\Psi_{N}\psi}{\partial x}
+\frac{\partial^{2}\psi}{\partial x^{2}}
+\frac{\gamma}{2}\frac{\partial{\cal H}[\psi]}{\partial x},
\qquad x\in\bfm R.
$$
The eigenvalue problem ${\cal A^{\prime}}_{N}v=\lambda v$
was studied in \cite{Vaynblat-Matalon00a}.
Obviously, eigenvalues of ${\cal A}_{N}$ and
${\cal A^{\prime}}_{N}$ are the same and the eigenfunctions
of the latter one are just $x$-derivatives of the eigenfunctions
of the operator ${\cal A}_{N}$.

In accordance with \cite{Vaynblat-Matalon00a}, operator
${\cal A^{\prime}}_{N}$ has a zero eigenvalue associated
with the $x$-shift invariance of \eq{1a}. The same is true for
${\cal A}_{N}$ as well. Moreover, zero is at least a double
eigenvalue of ${\cal A}_{N}$, because \eq{1a} is also
$\Phi$-shift invariant. Here, $x$- and $\Phi$-shift invariance
means, that if $\Phi(x,t)$ is a solution to \eq{1a}, then,
for any $C_{1},C_{2}\in\bfm R$, function $\Phi(x+C_{1},t)+C_{2}$
is its solution either.

If only solutions with the period $L$
are of interest, then they can be represented by the
Fourier series
$\phi(x,t)=\sum\limits_{k=-\infty}^{\infty}\widetilde{\phi}_{k}
e^{i2\pi kx/L}$. Substituting these
series into \eq{2_1c}, \eq{2_1d} multiplying the result
by $e^{i2\pi nx/L}$, and integrating over the interval
$x\in[-L/2,L/2]$, we obtain
\bequ
\frac{d\widetilde{\phi}_{k}}{dt}
=\left(-\frac{4\pi^{2}}{L^{2}}k^{2}
+\frac{\pi\gamma}{L}|k|\right)\widetilde{\phi}_{k}(t)
+i\frac{2\pi}{L}\sum\limits_{m=-\infty}^{\infty}m
\widetilde{\Psi_{N}}(k-m)\widetilde{\phi}_{m}(t),
\eequ{2_2a}
where $\widetilde{\Psi_{N}}(k)=L^{-1}\int_{-L/2}^{L/2}
\Psi_{N}(x)e^{-i2\pi kx/L}dx$ and $|k|<\infty$.
The integral can be written as
a linear combination of the integrals of type
$\int_{0}^{\pi}\cos my\ (\alpha-\cos y)^{-1}dy$ and
the latter one was evaluated by using entry 2.5.16.33,
p. 415 of \cite{PBM1} yielding
\bequ
\widetilde{\Psi_{N}}(k)=-i4\pi L^{-1}\sign(k)
e^{-i2\pi ka/L}\sum\limits_{n=1}^{N}e^{-2\pi b_{n}|k|/L},
\eequ{2_2d}

Introducing the representation of \eq{2_1c}
in the Fourier space $d\widetilde{\phi}/dt
=\widetilde{{\cal A}_{N}}\widetilde{\phi}$,
the Fourier image $\widetilde{{\cal A}_{N}}$
of the operator ${\cal A}_{N}$ is defined by the
$(k,m)$-th entry of its double
infinite ($-\infty<k,m<\infty$) matrix as follows
$$
(\widetilde{{\cal A}_{N}})_{k,m}
=\left(-4\pi^{2}L^{-2}k^{2}
+\pi\gamma L^{-1}|k|\right)\delta_{k,m}
$$
\bequ
+8\pi^{2}L^{-2}m\sign(k-m)e^{-i2\pi(k-m)a/L}
\sum\limits_{n=1}^{N}e^{-2\pi b_{n}|k-m|/L},
\qquad |k|,|m|<\infty,
\eequ{2_2e}
where $\delta_{k,m}$ is the Kronecker's symbol.

It can be shown, that the value of the free parameter $a$
does not affect neither spectral properties of
$\widetilde{{\cal A}_{N}}$ nor its $2$-norms.
Hence, we consider the case $a=0$ only.

\section{Pseudospectra of the linear operator}

\subsection{Computational techniques}\label{comp_tech}

In what follows we will work with matrix \eq{2_2e}
cut off at $|k|,|m|=K$, i.e. all $(k,m)$
entries of $\widetilde{{\cal A}_{N}}$
with either $|k|$ or $|m|$ greater than $K$ are neglected.
Thus, instead of matrix $\widetilde{{\cal A}_{N}}$
acting on double infinite vectors $\widetilde{\phi}$,
we consider the $(2K+1)\times(2K+1)$ matrix
$\widetilde{{\cal A}_{N}^{(K)}}$, whose entries
coincide with those of $\widetilde{{\cal A}_{N}}$
for $-K\le k,n\le K$.

In accordance with \cite{Trefethen99}, in order to estimate
possible nonmodal amplification of solutions in \eq{2_1c},
we first calculate values of 
$\left\|\left(z{\cal I}
-\widetilde{{\cal A}_{N}}^{(K)}\right)^{-1}\right\|_{2}$
as a function of the complex parameter $z$ for large enough
values of $K$. Level lines of this function form boundaries
of the pseudospectra of $\widetilde{{\cal A}_{N}}^{(K)}$,
see \cite{Trefethen97}. It was found in numerical experiments
that a good level of accuracy of the most interesting part
of the pseudospectra of $\widetilde{{\cal A}_{N}}$
is achieved if the cut off parameter $K$
is about $2L/\pi$ or greater. Thus, and in virtue of the
Parseval identity
$\left\|(z{\cal I}-{\cal A}_{N})^{-1}\right\|_{{\cal L}_{2}}$
$=\left\|\left({z\cal I}
-\widetilde{{\cal A}_{N}}\right)^{-1}\right\|_{2}$
, speaking about pseudospectra
or other ${\cal L}_{2}$-norm based functionals of
${\cal A}_{N}$ we actually mean those calculated
for $\widetilde{{\cal A}_{N}}^{(K)}$ with large
enough $K$. Here and in what follows $\cal I$ is the unity
operator or matrix of an appropriate size.

Calculations of 
$\left\|\left(z{\cal I}
-\widetilde{{\cal A}_{N}}^{(K)}\right)^{-1}\right\|_{2}$
can be carried out straightforwardly, however computational
costs can be reduced if
$\widetilde{{\cal A}_{N}}^{(K)}$ is transformed
appropriately.
By construction, matrix $\widetilde{{\cal A}_{N}}^{(K)}$
acts on vectors $\widetilde{u}^{(K)}$ $=(\widetilde{u}_{-K},
\ldots,\widetilde{u}_{-1},$ $\widetilde{u}_{0},$ $\widetilde{u}_{1},
\ldots,\widetilde{u}_{K})^{\top}$. Let us rearrange their
components and consider $\widetilde{w}^{(K)}$ 
$=\left[\widetilde{u}_{0},\right.$ 
$\left(\widetilde{u}^{(K)}_{-}\right)^{\top},$ 
$\left.\left(\widetilde{u}^{(K)}_{+}\right)^{\top}\right]^{\top}$,
where $\widetilde{u}^{(K)}_{\pm}$ 
$=(\widetilde{u}_{\pm 1},\ldots,\widetilde{u}_{\pm K})^{\top}$.
The permutation matrix
$\cal P$, corresponding to the proposed rearrangement
$\widetilde{w}^{(K)}={\cal P}\widetilde{u}^{(K)}$,
transforms $\widetilde{{\cal A}_{N}}^{(K)}$ into
${\cal P}\widetilde{{\cal A}_{N}}^{(K)}{\cal P}^{-1}$,
which acts on $\widetilde{w}^{(K)}$ and has the
following structure
\bequ
{\cal P}\widetilde{{\cal A}_{N}}^{(K)}{\cal P}^{-1}
=\left(\begin{array}{rrr}
0 & {\cal C} & {\cal C}\\
0 & {\cal D}+{\cal A}^{(1)} & {\cal A}^{(2)}\\
0 & {\cal A}^{(2)} & {\cal D}+{\cal A}^{(1)}\\
\end{array}\right).
\eequ{3_1a}
One zero eigenvalue of $\widetilde{{\cal A}_{N}}^{(K)}$,
corresponding to the $\Phi$-shift invariance of
\eq{1a}, can be seen from \eq{3_1a} explicitly.
For other blocks of \eq{3_1a} we have:
$$
\begin{array}{lll}
{\cal C}_{m} & =-8\pi^{2}L^{-2}|m|
\sum\limits_{n=1}^{N}e^{-2\pi b_{n}|m|/L}, &
m=1,2,\ldots,K,
\end{array}
$$

$$
\begin{array}{lll}
{\cal D}_{k,m} & =\left(-4\pi^{2}L^{-2}k^{2}
+\pi\gamma L^{-1}|k|\right)\delta_{k,m}, &
k,m=1,2,\ldots,K,\\
 & & \\
{\cal A}^{(1)}_{k,m} & =8\pi^{2}L^{-2}m\sign(k-m)
\sum\limits_{n=1}^{N}e^{-2\pi b_{n}|k-m|/L}, &
k,m=1,2,\ldots,K,\\
 & & \\
{\cal A}^{(2)}_{k,m} & =-8\pi^{2}L^{-2}m\sign(k+m)
\sum\limits_{n=1}^{N}e^{-2\pi b_{n}|k+m|/L}, &
k,m=1,2,\ldots,K.
\end{array}
$$

Following the idea of \cite{Vaynblat-Matalon00a}, we apply
the similarity transform
$$
{\cal T}=\frac{1}{\sqrt{2}}\left(\begin{array}{rrr}
1 & 0 & 0\\
0 & {\cal I} & {\cal I}\\ 0 & {\cal I} & -{\cal I}\\
\end{array}\right)
$$
to ${\cal P}\widetilde{{\cal A}_{N}}^{(K)}{\cal P}^{-1}$.
Here $\cal I$ is the unity $K\times K$ matrix corresponding
to the block structure of \eq{3_1a}. Unlike
\cite{Vaynblat-Matalon00a}, the normalizing coefficient
$1/\sqrt{2}$ was chosen to preserve the $2$-norm. The
transformed matrix 
${\cal T}{\cal P}\widetilde{{\cal A}_{N}}^{(K)}
({\cal T}{\cal P})^{-1}$
is decoupled into two diagonal blocks
\bequ
{\cal T}{\cal P}\widetilde{{\cal A}_{N}}^{(K)}
({\cal T}{\cal P})^{-1}
=2\left(\begin{array}{rrr}
0 & {\cal C} & 0\\
0 & {\cal D}+{\cal A}^{(1)}+{\cal A}^{(2)} & 0\\
0 & 0 & {\cal D}+{\cal A}^{(1)}-{\cal A}^{(2)}\\
\end{array}\right),
\eequ{3_1c}
and has the same $2$-norm as $\widetilde{{\cal A}_{N}}^{(K)}$.
The $2$-norm of \eq{3_1c} is the maximum of 2-norms of
its two blocks, each of which is of twice smaller size
than ${\cal T}{\cal P}\widetilde{{\cal A}_{N}}^{(K)}
({\cal T}{\cal P})^{-1}$.
In practice, the number of arithmetic operations
required to estimate the $2$-norm of a matrix is of
the order of the cube of its size. Therefore, estimation
of the $2$-norm of
$\widetilde{{\cal A}_{N}}^{(K)}$ through blocks of
${\cal T}{\cal P}\widetilde{{\cal A}_{N}}^{(K)}
({\cal T}{\cal P})^{-1}$
is more efficient. From our experience, the $2$-norms
of the blocks are of the same order of magnitude,
although the $2$-norm of the upper block supersedes
the lower one for most of practically important
values of $z$.

A straightforward and reliable way to calculate the $2$-norm
of the resolvent of $\widetilde{{\cal A}_{N}}^{(K)}$
(or of diagonal blocks of \eq{3_1c})
is through the singular value decomposition (SVD). Namely,
the reciprocal to the smallest singular value $s_{0}$ of
$z{\cal I}-\widetilde{{\cal A}_{N}}^{(K)}$ is equal to
$\left\|\left(z{\cal I}
-\widetilde{{\cal A}_{N}}^{(K)}\right)^{-1}\right\|_{2}$,
see \cite{Trefethen99}. The direct Matlab implementation of
SVD worked well in our case, though a few inverse iterations
with continuation in $z$, suggested in \cite{Lui97},
appeared to be as accurate and, on average, about
six times faster.

An alternative algorithm is
based on projection to the interesting subspace through
the Schur factorization followed by the Lanczos iterations.
It was suggested in \cite{Trefethen99} in the form of a
Matlab script
and is, on average, about two times faster than inverse
iterations with continuation. Further, our tests have
shown that its efficiency
degrades much slower as the matrix size or required
accuracy grows. Thus, Schur factorization with Lanczos
iterations was the algorithm of our choice. It was
intensively monitored by the direct SVD, however.

\begin{figure}[ht]

\centerline{\epsfig{figure=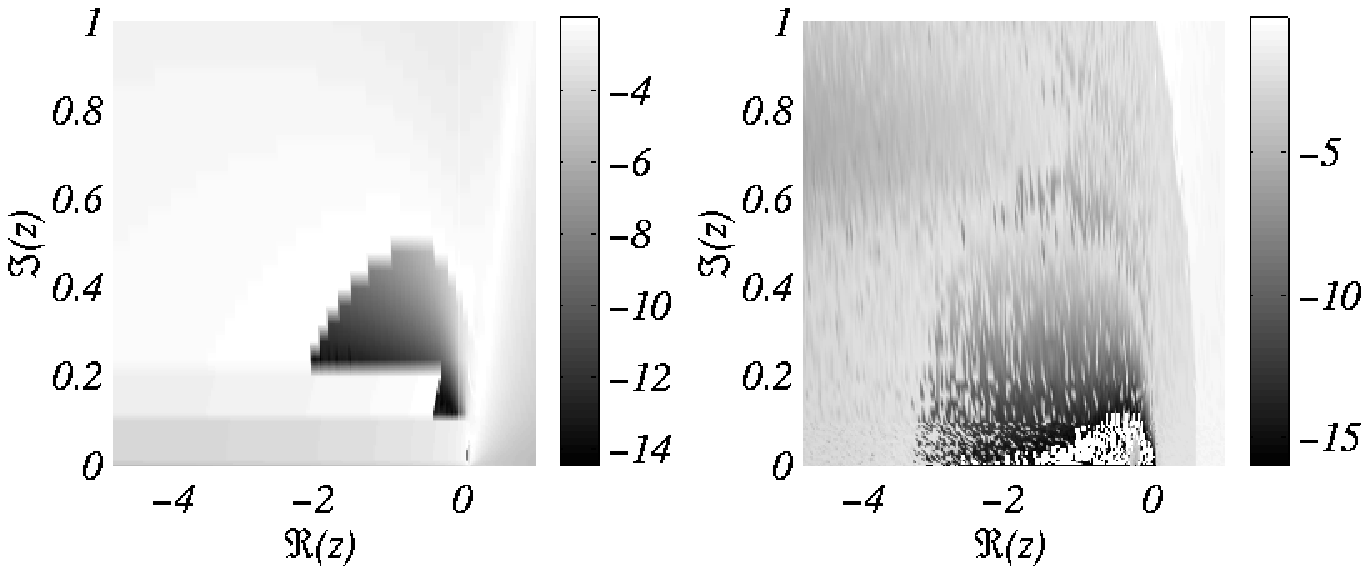,height=60mm,width=140mm}}

\centerline{\epsfig{figure=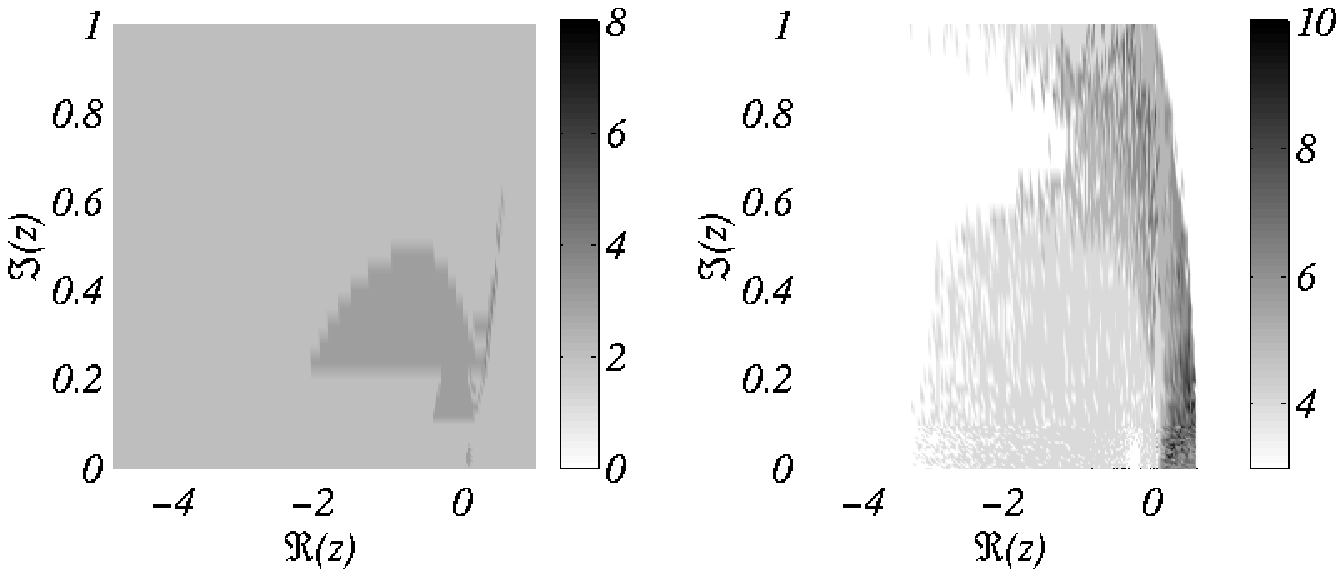,height=60mm,width=140mm}}

\caption{Relative increment of the smallest singular value
in the last iteration (top) and number of iterations (bottom)
carried out by methods \cite{Lui97} (left) and
\cite{Trefethen99} (right).}
\label{accuracy}
\end{figure}

A comparison of performance of the inverse iterations
with continuation and of the Schur factorization
with Lanczos iterations is given in Fig. \ref{accuracy}
for calculations of
$\left\|\left(z{\cal I}
-\widetilde{{\cal A}_{N}}^{(K)}\right)^{-1}\right\|_{2}$
with $L=40\pi$, $\gamma=0.8$ and $K=80$. The criteria of
stopping the iterations was
$\left|s_{0}^{(n)}-s_{0}^{(n-1)}\right|/s_{0}^{(n)}<0.01$,
i.e. when the relative increment of the $n$-th approximation
$s_{0}^{(n)}$ to the smallest singular value $s_{0}$ of
$z{\cal I}-\widetilde{{\cal A}_{N}}^{(K)}$ is
smaller than $\varepsilon=0.01$. Graphs reveal areas with
slower convergence of iterations.
Unlike the number of required inverse iterations
is usually less than the number of the Lanczos ones,
the latter are much cheaper computationally, resulting
in a significantly better overall performance.

\subsection{Structure of the pseudospectra}

As we are interested in stability of the steady coalescent
pole solutions and possible rate of linear growth of their
perturbations, the vicinity of the imaginary axis is of
principle interest. Reflection symmetry of the function
$\left\|\left(z{\cal I}-{\cal A}_{N}\right)^{-1}\right\|_{2}$
in regard to the real axis proves it sufficient, for our
purposes, to
study it in the region $z\in\{z:-5<\Re(z)<1,\ 0<\Im(z)<1\}$
only.

\begin{figure}[p]

\centerline{\epsfig{figure=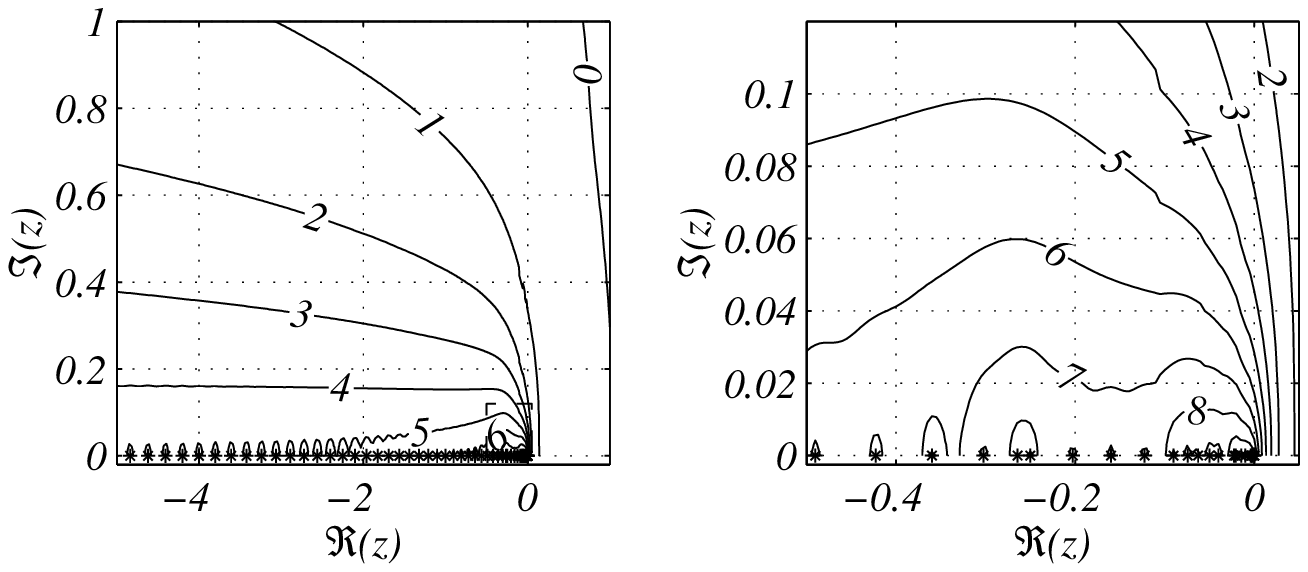,height=60mm,width=140mm}}
\vspace{-6.5mm}\hspace{10mm}(a)\hspace{70mm}(b)

\centerline{\epsfig{figure=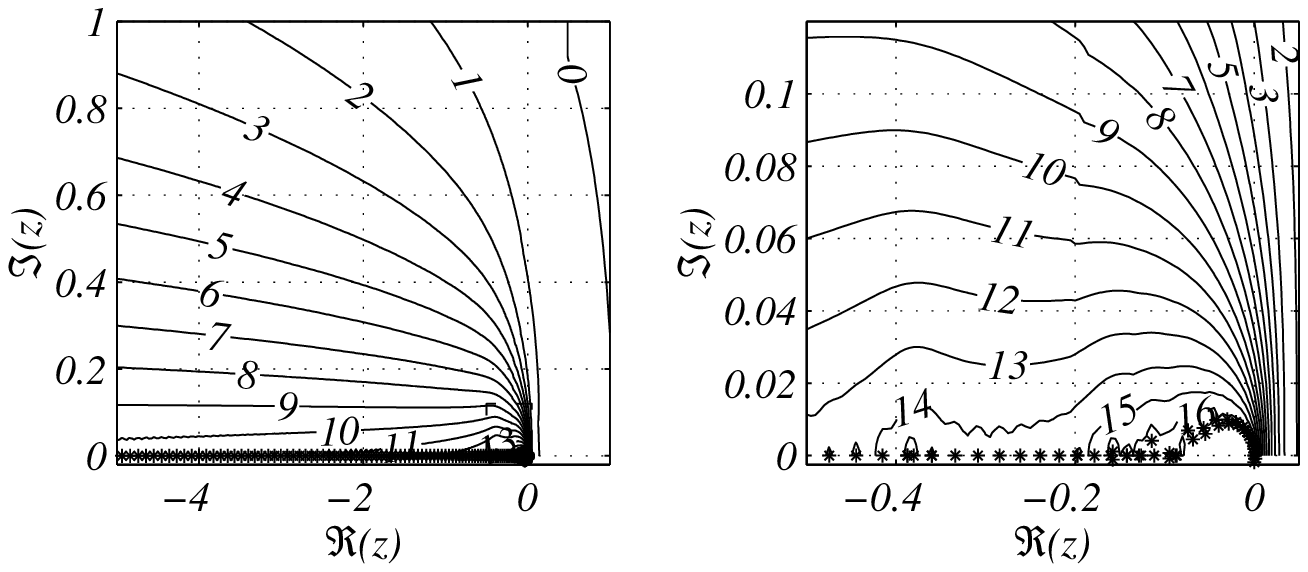,height=60mm,width=140mm}}
\vspace{-6.5mm}\hspace{10mm}(c)\hspace{70mm}(d)

\centerline{\epsfig{figure=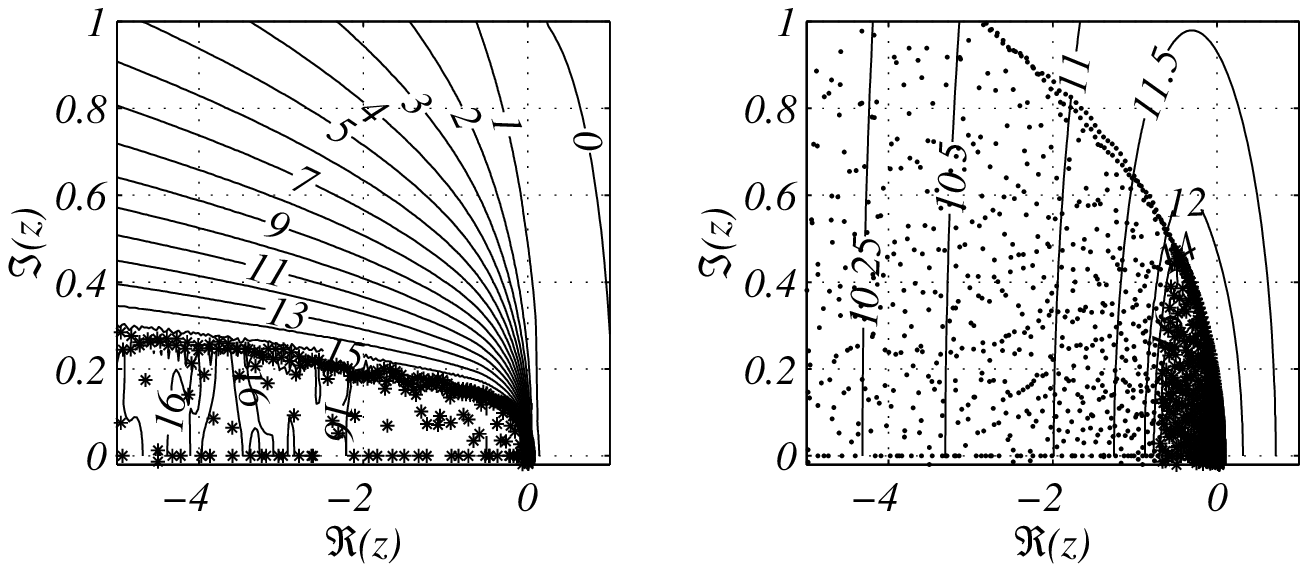,height=60mm,width=140mm}}
\vspace{-6.5mm}\hspace{10mm}(e)\hspace{70mm}(f)

\caption{Level lines of $\log_{10}\left\|\left(z{\cal I}
-{\cal A}_{N}\right)^{-1}\right\|_{2}$ for $\gamma=0.8$.
$L=40\pi$, $K=80$ in (a), (b); $L=90\pi$, $K=180$ in (c), (d);
$L=200\pi$, $K=400$ in (e); $L=1000\pi$, $K=2000$ in (f).}
\label{ps1_L40}
\end{figure}

Figures \ref{ps1_L40}a and \ref{ps1_L40}b illustrate
level lines of
$\left\|\left(z{\cal I}-{\cal A}_{N}\right)^{-1}\right\|_{2}$
for $L=40\pi$ and $\gamma=0.8$. A rectangle plotted
with a dashed line in Fig. \ref{ps1_L40}a marks the
location of the area magnified in Fig. \ref{ps1_L40}b.
Asterisks in the figures show approximations to the
eigenvalues of the operator ${\cal A}_{N}$.
These parameters correspond to the appearance of microcusps
in our direct numerical simulations with single accuracy.
The picture
suggests that the large area of high values of
$\left\|\left(z{\cal I}-{\cal A}_{N}\right)^{-1}\right\|_{2}$
near the origin can be the reason of significant
amplification of the round-off errors, which in the case
of single accuracy are of order $10^{-7}$.

Second critical case corresponding to the appearance
of microcusps in calculations with double accuracy,
see Fig. \ref{micro_cusps}, is
shown in Figs. \ref{ps1_L40}c and \ref{ps1_L40}d. Again,
a large region of huge values of
$\left\|\left(z{\cal I}-{\cal A}_{N}\right)^{-1}\right\|_{2}$
near $z=0$ suggests a possible match with the magnitude
of the round-off errors which are of order $10^{-16}$ in this
case.

Data on the $2$-norm of the resolvent of ${\cal A}_{N}$
for $L=200\pi$ and $\gamma=0.8$ are given in
Fig. \ref{ps1_L40}e. The figure shows
further widening of the area of large values of
$\left\|\left(z{\cal I}-{\cal A}_{N}\right)^{-1}\right\|_{2}$
near the real axis. Accordingly, calculated eigenvalues
spread further from the real axis and to the right from
the imaginary axis. Also, they tend to form a cluster
near the level line
$\left\|\left(z{\cal I}-{\cal A}_{N}\right)^{-1}\right\|_{2}$
$=10^{15}$, cf \cite{Trefethen99}.

Also, Fig. \ref{ps1_L40}e demonstrates that our
calculations fail to produce reliable results if
the smallest singular value $s_{0}$ of 
$z{\cal I}-\widetilde{{\cal A}_{N}}^{(K)}$ is less
than about $10^{-15}$. This is because of the effect
of the round-off errors of the computer on the
computational algorithm used to estimate $s_{0}$.
We see, however, that level lines of
$\left\|\left(z{\cal I}-{\cal A}_{N}\right)^{-1}\right\|_{2}$
corresponding to $s_{0}\ge 10^{-15}$
are much less sensitive to these round-off errors than the
eigenvalues.

All the algorithms for estimation of pseudospectra
mentioned in Section \ref{comp_tech} are subject to
the effect of the round-off errors and special arrangements
are required in order to get reliable results for
$s_{0}<10^{-15}$. In particular, calculations with
128-bit arithmetic, implemented in some computer systems,
can be used. However, for our purposes knowledge of
$\left\|\left(z{\cal I}-{\cal A}_{N}\right)^{-1}\right\|_{2}$
corresponding to $s_{0}\ge 10^{-15}$ is sufficient.

The last example of the pseudospectra for $L=1000\pi$
shown in Fig. \ref{ps1_L40}f was calculated in a different
way. The matrix $\widetilde{{\cal A}_{N}}^{(K)}$ for
$K=2000$ was projected into its eigenspace spanning
1000 eigenvectors corresponding to the eigenvalues with
the smallest absolute values. Then, the projected matrix
was used to estimate the level lines of 
$\left\|\left(z{\cal I}-{\cal A}_{N}\right)^{-1}\right\|_{2}$
depicted in Fig. \ref{ps1_L40}f. The eigenvalues used
for the projection are denoted by asterisks, the
neglected ones by dots. The eigenvalue problem for the
original matrix of size $4001\times 4001$ was solved
by the Matlab implementation of $QR$-iterations. We also
tried to apply Arnoldi iterations in accordance with
\cite{Wright-Trefethen01}, but could not make them
convergent even for the projection subspaces of smaller
dimensions and for smaller values of $L$. 

The direct calculation of the approximation to the
spectrum of ${\cal A}_{N}$, namely eigenvalues of
$\widetilde{{\cal A}_{N}}^{(K)}$ presented in Figs.
\ref{ps1_L40}a - \ref{ps1_L40}f with asterisks and dots,
was undertaken by the Matlab implementation of $QR$-iterations.
In the case of $L=90\pi$ six directly calculated eigenvalues
are located to the right from the imaginary axis,
see also \cite{Olami-Galanti-Kupervasser-Procaccia97}.
The number of eigenvalues in the right half of the
complex plane grows for larger $L$. However, the pseudospectra
plotted in Figs. \ref{ps1_L40}d - \ref{ps1_L40}f suggest
that these unstable eigenvalues cannot be
trusted. They appear in the vast area of large values of
$\left\|\left(z{\cal I}-{\cal A}_{N}\right)^{-1}\right\|_{2}$
and, in accordance with e.g. \cite{Trefethen97}, can
be very sensitive to the perturbations as small as
$10^{-16}$, which is on the level of the machine zero in
the case in question. 

\begin{figure}[ht]

\centerline{\epsfig{figure=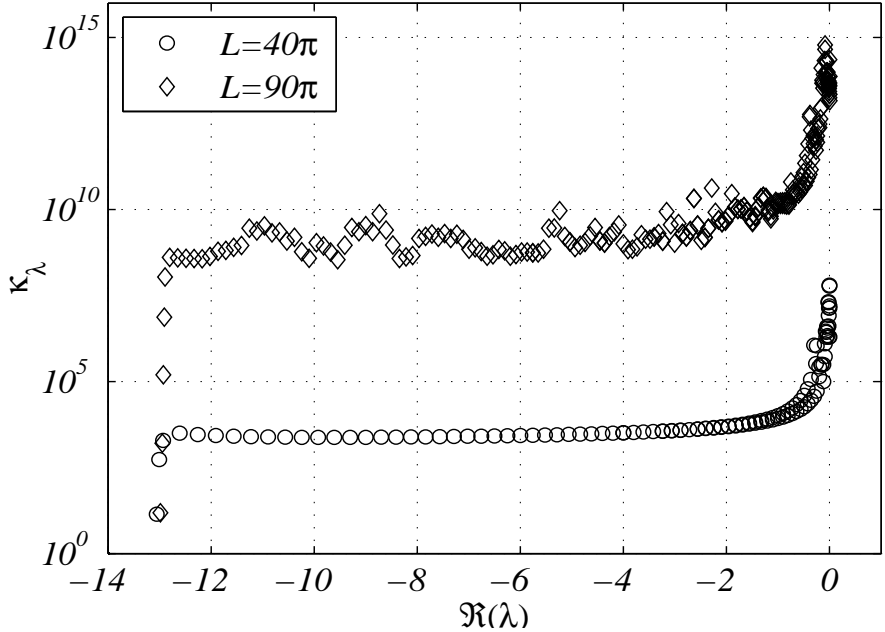,height=85mm,width=120mm}}

\caption{Condition numbers of the eigenvalues $\lambda$ of
$\widetilde{{\cal A}_{N}}^{(K)}$ for $\gamma=0.8$.}
\label{eig_cond}
\end{figure}

Every eigenvalue $\lambda$ of $\widetilde{{\cal A}_{N}}^{(K)}$
can be associated with a condition number $\kappa_{\lambda}$
$=|\sum_{i}w_{i}u_{i}|^{-1}$, where $w$ and $u$ are corresponding
normalized left and right eigenvectors of
$\widetilde{{\cal A}_{N}}^{(K)}$, see \cite{Wilkinson65}.
Then, eigenvalues
of $\widetilde{{\cal A}_{N}}^{(K)}$ will be perturbed
by $\kappa_{\lambda}\|{\cal E}\|_{2}$ at most, if
$\widetilde{{\cal A}_{N}}^{(K)}$ is perturbed by 
matrix $\cal E$ with small enough $\|{\cal E}\|_{2}$.
Figure \ref{eig_cond} illustrates these condition numbers
for $L=40\pi$ and $90\pi$, making a very good match to
the magnitude of perturbations of eigenvalues given in
Fig. \ref{ps1_L40}d. Note, the rightmost eigenvalues are
worst conditioned.

We would like to stress that because of the severe
nonnormality of $\widetilde{{\cal A}_{N}}^{(K)}$ some
of its directly calculated eigenvalues may have nothing
in common with what they should be in absence of the
round-off errors. A particular numerical method can
even worsen the estimation indeed. However, no one
method can reduce the perturbation associated with the
approximation of entries of
$\widetilde{{\cal A}_{N}}^{(K)}$ by the finite-digit
arithmetic of the computer, cf \cite{Vaynblat-Matalon00a}.
The only way to increase the accuracy of the direct
eigenvalue computations for $L\ge 90\pi$ is to use
a more accurate computer arithmetic with machine
zero smaller than the reciprocal of
$\max\limits_{\lambda\in\Lambda(\widetilde{{\cal A}_{N}}^{(K)})}
\{\kappa_{\lambda}\}$, where
$\Lambda(\widetilde{{\cal A}_{N}}^{(K)})$ is the
spectrum of $\widetilde{{\cal A}_{N}}^{(K)}$.

\section{Estimation of the transient amplification}

\subsection{Kreiss constants}\label{lower}

A robust lower bound on
$\left\|e^{t{\cal A}_{N}}\right\|_{{\cal L}_{2}}$
can be obtained from the Laplace transform of
$e^{t{\cal A}_{N}}$, which under certain conditions
(see \cite{Pazy85}) can be written as
$$
\int\limits_{0}^{\infty}e^{-zt}e^{t{\cal A}_{N}}dt
=(z{\cal I}-{\cal A}_{N})^{-1}.
$$
Considering norms of both
sides of this relation and carrying out straightforward
estimations of the integral:
$$
\|(z{\cal I}-{\cal A}_{N})^{-1}\|_{{\cal L}_2}
=\left\|\int\limits_{0}^{\infty}e^{-zt}e^{t{\cal A}_{N}}dt\right\|_{{\cal L}_2}
\le\sup\limits_{t>0}\|e^{t{\cal A}_{N}}\|_{{\cal L}_2}
\int\limits_{0}^{\infty}e^{-\Re(z)t}dt,
$$
we arrive at
$\sup\limits_{t>0}\|e^{t{\cal A}_{N}}\|_{{\cal L}_2}
\ge\Re(z)\|(z{\cal I}-{\cal A}_{N})^{-1}\|_{{\cal L}_2}$.
The latter is valid for all $z$ with positive real part
yielding
\bequ
\sup\limits_{t>0}\|e^{t{\cal A}_{N}}\|_{{\cal L}_2}
\ge\sup\limits_{\Re(z)>0}
\left[\Re(z)\|(z{\cal I}-{\cal A}_{N})^{-1}\|_{{\cal L}_2}\right]
={\cal K}_{{\cal A}_{N}},
\eequ{4_1a}
where ${\cal K}_{{\cal A}_{N}}$ is called the Kreiss
constant, see also \cite{Reddy-Schmid-Henningson93},
\cite{Trefethen97}. In other words, if ${\cal K}_{{\cal A}_{N}}$ is
the Kreiss constant of the operator ${\cal A}_{N}$,
then there is a perturbation $\phi_{*}(x,t)$ governed by \eq{2_1c}
and a time instance $t_{*}$ such that the initial value
of $\phi_{*}(x,t)$ is amplified at least ${\cal K}_{{\cal A}_{N}}$
times in terms of its ${\cal L}_{2}$ norm, i.e.
$\|\phi_{*}(x,t_{*})\|_{{\cal L}_{2}}$
$\ge{\cal K}_{{\cal A}_{N}}\|\phi_{*}(x,0)\|_{{\cal L}_{2}}$.

\begin{figure}[ht]

\centerline{\epsfig{figure=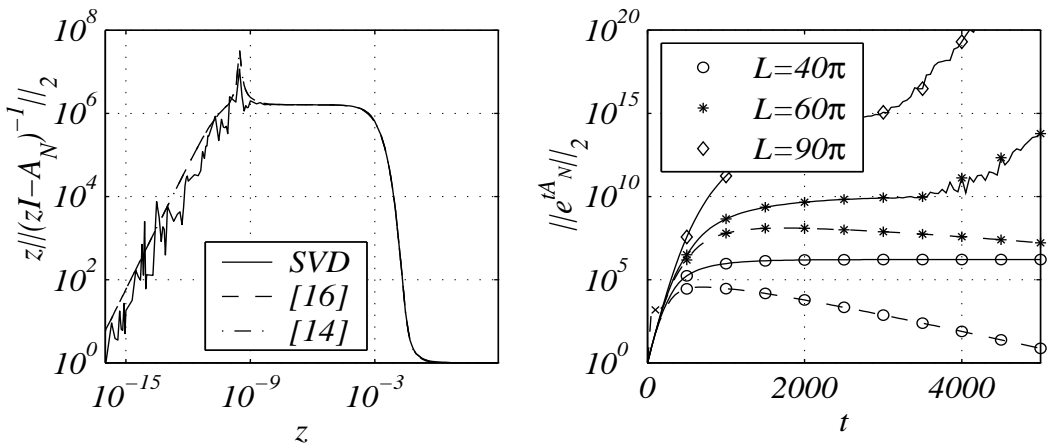,height=60mm,width=140mm}}

\caption{An approximation to $z\left\|\left(z{\cal I}
-{\cal A}_{N}\right)^{-1}\right\|_{2}$
versus $z$ on the real axis for $L=40\pi$
and $K=160$ (left). Dependence of ${\cal L}_{2}$ norms of
$C_{0}$-semigroups generated by 
$\widetilde{{\cal A}_{N}}^{(K)}$ (solid lines) and
$\widetilde{{\cal B}_{N}}^{(K)}$ (dashed lines) on $t$ (right).
Here $\gamma=0.8$.}
\label{Kreiss}
\end{figure}

Our studies of pseudospectra represented, in
particular, in Figs. \ref{ps1_L40}a - \ref{ps1_L40}f indicate
that the supremum in \eq{4_1a} is reached on the real axis.
Figure \ref{Kreiss} shows dependence of the function
$z\left\|\left({\cal I}
-z\widetilde{{\cal A}_{N}}^{(K)}\right)^{-1}\right\|_{2}$
on $z$ for $\Im(z)=0$. We depicted results obtained by
three different techniques and they are in good
agreement with each other except for very small $z$.
The discrepancy for $z\le 10^{-9}$ is because of
the round-off errors as explained in the previous
section. The smallest singular value $s_{0}$ of
${\cal I}-z\widetilde{{\cal A}_{N}}^{(K)}$
should be of order $10^{-15}$ to result in
$z\left\|\left({\cal I}
-z\widetilde{{\cal A}_{N}}^{(K)}\right)^{-1}\right\|_{2}$
$\approx 10^{6}$ for $z\approx 10^{-9}$. Indeed, this
value of $s_{0}$
is too small to be accurately calculated on a computer
with machine zero of order $10^{-16}$. It is quite
reliable to conclude in this case that
${\cal K}_{{\cal A}_{N}}\ge 1.6\times 10^{6}$
for $L=40\pi$ and $\gamma=0.8$.

Because of the effect of the round-off errors on the
computation of $s_{0}$, similar estimations of the
Kreiss constant for $L\ge 80\pi$ on a computer with
the machine zero of order $10^{-16}$ are not accurate
yielding a saturated value of order $10^{13}$. Instead,
we have calculated more values of the Kreiss constant
for a set of smaller $L$. Results are presented in
Table \ref{Tab_Kreiss} and Fig. \ref{d1_Joulin}.

\begin{table}[ht]
\caption{Estimated Kreiss constants of ${\cal A}_{N_{L}}$}
\centerline{
\begin{tabular}{|l|l|l|l|l|l|}
\hline
$L/\pi$ & $10$ & $20$ & $30$ & $40$ & $50$\\
${\cal K}_{{\cal A}_{N}}$ & $7.0\times 10^{0}$ &
 $3.3\times 10^{2}$ & $2.1\times 10^{4}$ &
 $1.6\times 10^{6}$ & $1.3\times 10^{8}$\\
\hline
$L/\pi$ & $60$ & $70$ & $80$ & $90$ & $100$ \\
${\cal K}_{{\cal A}_{N}}$ & $1.2\times 10^{10}$ &
$4.0\times 10^{11}$ & $5.0\times 10^{12}$ &
 $1.5\times 10^{13}$ & $1.5\times 10^{13}$\\
\hline
\end{tabular}
}
\label{Tab_Kreiss}
\end{table}

\subsection{Norms of the $C_{0}$-semigroup}

Good supplementary proof of essential nonmodal
amplification can be provided by direct estimation
of the ${\cal L}_{2}$ norm of the $C_{0}$-semigroup
generated by ${\cal A}_{N}$. Similar to the
previous estimations, we have calculated the $2$-norm
of the $C_{0}$-semigroup generated by
$\widetilde{{\cal A}_{N}}^{(K)}$. Matlab's
implementation of a scaling and squaring algorithm
with a Pad\'{e} approximation has been used in order
to avoid calculation of the Jordan decomposition of
$\widetilde{{\cal A}_{N}}^{(K)}$. The results
revealed a good convergence for $K\ge 2L/\pi$.

As we have mentioned in Section \ref{nillspace},
operator ${\cal A}_{N}$ has a nontrivial null-space
${\cal N}({\cal A}_{N})$. Because of this
$\left\|e^{t{\cal A}_{N}}\right\|_{{\cal L}_{2}}$
does not decay for $t\rightarrow\infty$ and, moreover,
it grows slowly because of the round-off errors.
In order to remove the effect of the null-space on
the asymptotics of decay, and also, to demonstrate
that the amplification observed in numerical experiments
was not caused by that double zero eigenvalue, associated
with the translational modes, we have projected
$\widetilde{{\cal A}_{N}}^{(K)}$ into its eigenspace
${\cal N}\left(\widetilde{{\cal A}_{N}}^{(K)}\right)^{\perp}$
orthogonal to
${\cal N}\left(\widetilde{{\cal A}_{N}}^{(K)}\right)$.
The 2-norms of the $C_{0}$-semigroups generated by the
resulting operator, denoted here as
$\widetilde{{\cal B}_{N}}^{(K)}$, are depicted in Fig.
\ref{Kreiss}b alongside with the similar data for the
original operator $\widetilde{{\cal A}_{N}}^{(K)}$.

Construction of $\widetilde{{\cal B}_{N}}^{(K)}$
for larger values of $L$ is complicated by difficulties
with the accurate identification of the eigenfunctions
corresponding to the zero eigenvalues. The latter ones
appear to be perturbed and are as distant from $z=0$
as a few other eigenvalues. Note, that projection of
$\widetilde{{\cal A}_{N}}^{(K)}$ into 
${\cal N}\left(\widetilde{{\cal A}_{N}}^{(K)}\right)^{\perp}$
affects the $C_{0}$-semigroup not only asymptotically
for $t\rightarrow\infty$, but for $t\rightarrow 0$ as well.

\begin{figure}[ht]

\centerline{\epsfig{figure=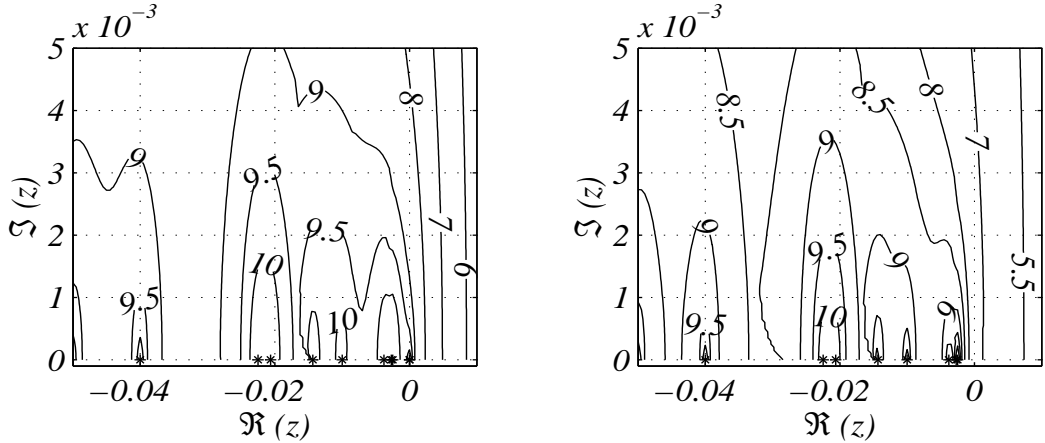,height=60mm,width=140mm}}

\caption{Comparison of level lines of
$\log_{10}\left\|\left(z{\cal I}
-\widetilde{{\cal A}_{N}}^{(K)}\right)^{-1}\right\|_{2}$
(left) and $\log_{10}\left\|\left(z{\cal I}
-\widetilde{{\cal B}_{N}}^{(K)}\right)^{-1}\right\|_{2}$
(right) for $L=40\pi$, $\gamma=0.8$, and $K=160$.}
\label{A_to_B_gr}
\end{figure}

Data in Fig. \ref{Kreiss}b matches our estimations of
the lower bound of the possible amplification of
perturbations to the Sivashinsky equation and their
extrapolations for larger values of $L$. Also,
they show that presence of the nontrivial null space,
corresponding to the shift invariance of the equation
is not responsible for high sensitivity of the steady
coalescent pole solutions to the noise. The latter
conclusion is reinforced by the comparison of the
pseudospectra of $\widetilde{{\cal A}_{N}}^{(K)}$ and
$\widetilde{{\cal B}_{N}}^{(K)}$. On scales of
Figs. \ref{ps1_L40}a and \ref{ps1_L40}b they are simply
indistinguishable and can only be seen in a very close
proximity of the origin, as shown in Fig. \ref{A_to_B_gr}.

One may see that the only effect of the projection is
a small shift of the pseudospectra to the left, resulting
in the reduction of the Kreiss constant of about 30 times.
It is still well above $10^{4}$, however, perfectly matching
the corresponding curve in Fig. \ref{Kreiss}b.

\subsection{Condition numbers}

A traditional estimation of the $C_{0}$-semigroup
generated by ${\cal A}_{N}$ is given by
\bequ
\exp\left\{t\inf\limits_{z\in\Lambda({\cal A}_{N})}[\Re(z)]\right\}
\le\left\|e^{t{\cal A}_{N}}\right\|_{{\cal L}_{2}}
\le\varkappa_{2}({\cal A}_{N})
\exp\left\{t\sup\limits_{z\in\Lambda({\cal A}_{N})}[\Re(z)]\right\},
\eequ{4_1c}
where $\Lambda({\cal A}_{N})$ is the spectrum
of ${\cal A}_{N}$, see \cite{Pazy85}.
If ${\cal A}_{N}$ is a finite-dimensional operator,
then $\varkappa_{2}({\cal A}_{N})$ is the condition
number $\varkappa_{2}({\cal A}_{N})=\cond_{2}(V)
=\|V\|_{2}\|V^{-1}\|_{2}$ of the matrix $V$ whose
columns are formed by the eigenvectors of ${\cal A}_{N}$.
For infinite-dimensional operators, meaning of
$\varkappa_{2}({\cal A}_{N})$ is not so
straightforward and, what is even more disappointing,
it is often infinitely large, see \cite{Trefethen97}.
However, we try to estimate $\kappa_{2}({\cal A}_{N})$,
because if successful it would give an estimation of
the upper bound of
$\left\|e^{t{\cal A}_{N}}\right\|_{{\cal L}_{2}}$
following from \eq{4_1c} for
$\sup\limits_{z\in\Lambda({\cal A}_{N})}[\Re(z)]=0$ as
follows:
\bequ
\left\|e^{t{\cal A}_{N}}\right\|_{{\cal L}_{2}}
\le\varkappa_{2}({\cal A}_{N}).
\eequ{4_3a}

\begin{figure}[ht]

\centerline{\epsfig{figure=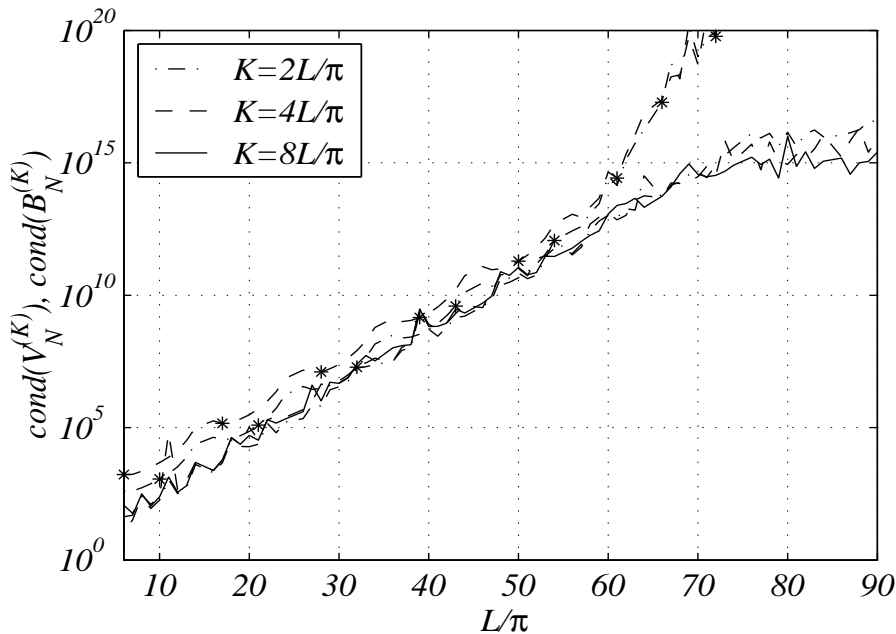,height=85mm,width=120mm}}

\caption{Dependence of condition numbers of $V_{N}^{(K)}$
and $\widetilde{{\cal B}_{N}}^{(K)}$ (marked with asterisks)
on $L$ for $\gamma=0.8$.}
\label{condU}
\end{figure}

Figure \ref{condU} depicts graphs of
$\kappa_{2}\left(\widetilde{{\cal A}_{N}}^{(K)}\right)$
$=\cond_{2}(V_{N}^{(K)})$
versus $L$ for different cut off parameters $K$. Here
columns of matrix $V_{N}^{(K)}$ are eigenvectors of matrix
$\widetilde{{\cal A}_{N}}^{(K)}$. The difference between
$\kappa_{2}\left(\widetilde{{\cal A}_{N}}^{(K)}\right)$
for different $K$ may look small on the graph. However,
the graph is in the $\log_{10}$ scale and the discrepancy
is on the level of an order of magnitude. Hence,
convergence is not obvious and we do not pose obtained
$\kappa_{2}\left(\widetilde{{\cal A}_{N}}^{(K)}\right)$
as an estimation of the upper bound of
$\left\|e^{t{\cal A}_{N}}\right\|_{{\cal L}_{2}}$
in \eq{4_3a}.

The graph of
$\cond_{2}\left(\widetilde{{\cal B}_{N}}^{(K)}\right)$
versus $L$ is also illustrated in Fig \ref{condU}.
Unlike $\kappa_{2}({\cal A}_{N})$, which estimates
the upper bound of amplification of solutions of the
initial-value problem for \eq{2_1c}, the number
$\cond_{2}\left(\widetilde{{\cal B}_{N}}^{(K)}\right)$
gives an estimation of possible
amplification of perturbations of the right hand side $f$
in the solution $u_{f}$ of the linear equation
$\widetilde{{\cal B}_{N}}^{(K)}u=f$. Note, because of
$x$- and $\Phi$-shift invariance of \eq{1a}, condition
number of $\widetilde{{\cal A}_{N}}^{(K)}$ itself is
infinite.

\section{Comparison of the estimations}\label{discuss}

It was established in numerical experiments, see e.g.
\cite{Rahibe-Aubry-Sivashinsky96},
\cite{Rahibe-Aubry-Sivashinsky98}, that for small enough
computational domains of size $L<L_{c}$ numerical
solutions of \eq{1a} stabilize to the steady
coalescent $N_{L}$-pole solutions of \eq{1a}.
This observation is in the explicit agreement with the
eigenvalue analysis of the linearized problem carried
out in \cite{Vaynblat-Matalon00a}.

For larger $L>L_{c}$, numerical solutions do
not stabilize to any steady solution at all. Instead,
being essentially nonsteady, they remain very closely
to the steady coalescent $N_{L}$-pole solution,
developing on the surface of the flame front small
cusps randomly in time. With time these small cusps
move towards the trough of the flame front profile
and disappear in it as can be seen in Fig.
\ref{micro_cusps}.

\begin{figure}[ht]
\centerline{\epsfig{figure=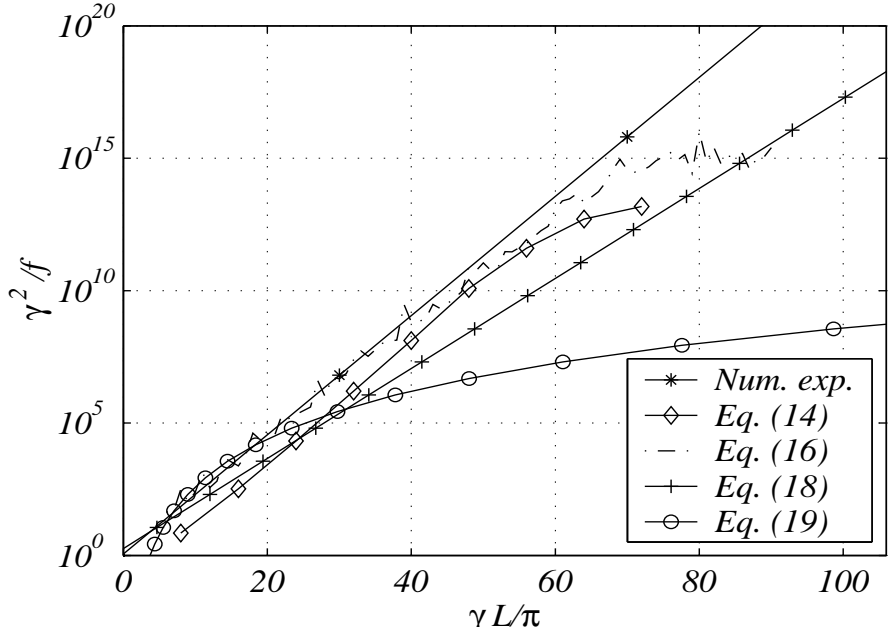,height=85mm,width=120mm}}
\caption{Dependence of the variety of measures of
the critical strength $f_{c}$ of perturbations on the
flame size $L$.}
\label{d1_Joulin}
\end{figure}

Numerous numerical experiments did not reveal any
significant dependence of the critical length $L_{c}$
on parameters of the computational algorithm. They
have shown, however, that $L_{c}$ is effectively affected
by the round-off errors \cite{Karlin02}.
Thus, if $f$ is the order of
the amplitude of perturbations associated with the
round-off errors, then $L_{c}=L_{c}(f)$. Two values
of $L_{c}(f)$ obtained in our calculations with
32- and 64-bit arithmetic are shown in Fig.
\ref{d1_Joulin}. Amplitude $f$ of the perturbations
was of the order of machine zeros, i.e. $10^{-7}$ and
$10^{-16}$ correspondingly. Note, that in calculations
with 32-bit arithmetic round-off errors dominated
discretization errors \cite{Karlin-Mazya-Schmidt02}.

It is convenient to invert the relation $L_{c}=L_{c}(f)$
and write it in the form $f_{c}=f_{c}(L)$, where
$f_{c}$ is a critical noise strength for given
size $L$ of the flame.
Reciprocal of the Kreiss constant
${\cal K}_{{\cal A}_{N}}$, obtained in
Section \ref{lower}, can be considered as the lower
bound of this critical strength $f_{c}$ of perturbations
for any particular value of $L$. Here, the strength of
the perturbation means its 2-norm. Corresponding
graph is plotted in Fig. \ref{d1_Joulin}. It is in
a very good agreement with the results of our direct
numerical simulations. The graph of
$\kappa_{2}\left(\widetilde{{\cal A}_{N}}^{(K)}\right)$
versus $L$ is also given in Fig. \ref{d1_Joulin},
for the illustrative purposes. We remind,
that there was no evidence of convergence of
$\kappa_{2}\left(\widetilde{{\cal A}_{N}}^{(K)}\right)$
to $\kappa_{2}\left({\cal A}_{N}\right)$ in
our calculations and interpretation of the graph
as the upper bound \eq{4_3a} of $f_{c}$ is not justified.

An analytical attempt to estimate the value of $f_{c}$ was
made in \cite{Joulin89b} where the following modification of
\eq{2_1c}, \eq{2_1d} has been considered:
\bequ
\frac{\partial\phi}{\partial t}
=\frac{x}{R_{N}}\frac{\partial\phi}{\partial x}
+\frac{\partial^{2}\phi}{\partial x^{2}}
+\frac{\gamma}{2}\frac{\partial{\cal H}[\phi]}{\partial x},
\qquad x\in\bfm R.
\eequ{5_1a}
Here $R_{N}=\left(\partial^{2}\Phi_{N}/\partial x^{2}\right)^{-1}$
is calculated in the crest of the steady coalescent
$N$-pole solution, see Fig. \ref{micro_cusps}.
In \cite{Joulin89b} a particular asymptotic solution
to \eq{5_1a} has been investigated. As a result,
the dependence between the critical value of curvature radius
$R_{N}$ in the crest of the flame profile and the
spectral density $\rho_{f}$ of the most dangerous
harmonics of $\phi(x,0)$ has been obtained. A functional link
between $L$ and $R_{N_{L}}$ of the steady coalescent
$N_{L}$-pole solution to the Sivashinsky equation
can be easily established yielding
\bequ
\rho_{f,c}=4^{-1}\gamma^{2}e^{-\gamma^{2}(c_{1}L+c_{2})/8}.
\eequ{5_1b}
Here $c_{1}$ and $c_{2}$ are coefficients of the least
squares fitting of $R_{N_{L}}=R_{N_{L}}(L)$ with a
straight line $c_{1}L+c_{2}$.

When comparing our results with estimation \eq{5_1b},
the following should be taken into account. First,
relation \eq{5_1b} has been obtained for the spectral
density of the most dangerous harmonics of the perturbation
$\phi(x,0)$ rather than for its amplitude $f$.
Second, assumptions made to obtain \eq{5_1b} are
better justified for large $L$.
The last but not least factor is that \eq{5_1b}
is based on a particular solution and is likely to
produce an overestimated value of $\rho_{f,c}$
rather than the optimal one. In view of these
peculiarities, the agreement between \eq{5_1b}, obtained
in \cite{Joulin89b}, and our estimations is striking.

In contrast, the estimation of $f_{c}$ obtained in
\cite{Olami-Galanti-Kupervasser-Procaccia97} is obviously
out of the harmony. That estimation was
based on studies of the dynamics of poles governed by
\eq{1c_a}, \eq{1c_b}. Namely, the amplitude of
perturbations to the solutions of \eq{1a} was
linked to the $b$-coordinate of poles in the
$(a,b)$-plane. Then, analysis of the dynamics of these
noise generated poles yields the estimation
\bequ
f_{c}=2^{11}\pi^{6}\gamma^{-5}L^{-6}.
\eequ{5_1c}

There is no doubt that the
sensitivity of system \eq{1c_a}, \eq{1c_b} to noise
is totally different of what we have for \eq{1a}.
Analysis of the Jacobian of the right hand sides of
system \eq{1c_a}, \eq{1c_b} for the steady coalescent
$N_{L}$-pole solution reveals that this is a symmetric
matrix and there is no linear nonmodal amplification
of noise in \eq{1c_a}, \eq{1c_b} at all. For small $L$,
when the nonmodal amplification is not essential,
estimation \eq{5_1c} is in a good agreement with other
data indeed. However, for larger $L$, the discrepancy
between the results of
\cite{Olami-Galanti-Kupervasser-Procaccia97} and of others,
clearly seen in Fig. \ref{d1_Joulin}, can be interpreted
as the measure of the importance of the linear nonmodal
amplification of perturbations in the Sivashinsky
equation.

\section{Conclusions}

In this paper we have undertaken the numerical analysis
of norms of the resolvent of the linear operator
associated with the Sivashinsky equation linearized in a
neighbourhood of the steady coalescent pole solutions.
Performance of available numerical techniques was
compared to each other and the results are checked
versus directly calculated norms of the evolution
operator.

The studies demonstrated the robustness of the approach
by resolving the problem of stability of certain types
of cellular flames. They showed that the round-off
errors are the only effect relevant to the appearance
of the micro cusps in computations of large enough
flames. These essentially nonlinear micro cusps are
generated through the huge linear nonmodal transitional
amplification of the round-off errors. Their final
appearance and dynamics on the flame surface is
governed by essentially nonlinear mechanisms intrinsic
to the Sivashinsky equation.

In order to retain its physical meaning for large
flames, Sivashinsky equation should be refined by
accounting for the physical noise, e.g. in the way
suggested in \cite{Cambray-Joulin92}.

\section{Acknowledgements}

This research was supported by the EPSRC research
grant GR/R66692.

\end{document}